\documentclass[aps,prb,twocolumn,floatfix,superscriptaddress]{revtex4-1}

\usepackage{graphicx,color}
\usepackage{amsfonts}
\usepackage[figuresright]{rotating}
\usepackage{amssymb}
\usepackage{amsmath}
\usepackage{pifont}
\usepackage{psfrag}
\usepackage{subfigure}
\usepackage{multirow}
\usepackage{tabularx}
\usepackage{textcomp}
\usepackage{units}
\usepackage{hyperref}
\usepackage{longtable}
\usepackage{textcomp}

\hypersetup{
 pdfnewwindow=true, colorlinks=true,
 linkcolor=blue, anchorcolor=blue,
 citecolor=blue, filecolor=blue,
 menucolor=blue, urlcolor=blue}

\def\nn{\nonumber}

\def\beq{\begin{eqnarray}}
\def\eeq{\end{eqnarray}}

\renewcommand{\v}[1]{\ensuremath{\mathbf{#1}}}   % for vectors

\let\baraccent=\= % rename builtin command \= to \baraccent
\renewcommand{\=}[1]{\stackrel{#1}{=}} % for putting numbers above =

\makeatletter

\begin{document}

\title{Terahertz radiation of jerk photocurrent}

\author{Bernardo S.\ \surname{Mendoza}}
\affiliation{Centro de Investigaciones en Optica, A.C., Leon, 37150 Guanajuato, Mexico}

\author{Benjamin\ \surname{M. Fregoso}}
\affiliation{Department of Physics, Kent State University, Kent, Ohio 44242, USA}

\begin{abstract}
We compute the jerk current tensor of GaAs, Si, and ferroelectric single-layer GeS, GeSe, SnS, and SnSe. We find peak values of the order of $10^{14}$ mA/V$^3$s$^2$ in GaAs and Si within the visible energy spectrum and an order of magnitude larger in single-layer GeS, GeSe, SnSe and SnS. We show that the detailed knowledge of this tensor and its large value in single-layer GeS, GeSe, SnSe and SnS make it possible to predict the magnitude and angle of rotation of polarization of intense terahertz pulses generated in photoconductive switches and point to alternative functionalities of these devices.
\end{abstract}

\maketitle

\section{Introduction}
Understanding and controlling light-matter interactions is at the forefront of scientific discovery and technological applications. This is especially true for nonlinear optical effects.  Second harmonic generation~\cite{Boyd2008} for example, is routinely used as a frequency multiplier, as a probe of material's crystal symmetry, as an efficient enabler of entangled photons in quantum protocols, etc. Recently, the bulk photovoltaic effect (BPVE)\cite{Sturman1992,Paillard2018,Rioux2012,Baltz1981,Sipe2000,Morimoto2016,Spanier2016,Titova2020,Kushnir2017,Cook2017,Fregoso2017,Rangel2017,Ibanez-Azpiroz2018,Panday2019,Nakamura2017,Ogawa2017,Nagaosa2017,Wang2017b,Nakamura2018,Gong2018,Kushnir2019,Sotome2019,Burger2019,Sotome2019a,Kral2000,Zhang2019,Brehm2018,Hosur2011,Chan2017,Juan2017,Rees,Flicker2018,Parker2019,Tan2019,Barik,Fregoso2018,Fregoso2019}, i.e., the generation of direct current (dc) in illuminated insulators lacking inversion symmetry, has attracted renewed attention for its potential optoelectronic application using topological insulators~\cite{Hosur2011}, novel two-dimensional (2D) ferroelectric materials~\cite{Titova2020,Kushnir2017,Cook2017,Fregoso2017,Rangel2017,Ibanez-Azpiroz2018,Panday2019} and Weyl semimetals~\cite{Chan2017,Juan2017,Rees}. Traditionally, the BPVE refers to a second order effect in the optical field. It was recently extended to higher orders, giving an explicit expression for the photoconductivity, i.e., intensity-dependent conductivity in terms of Bloch wave function of the crystal~\cite{Fregoso2018,Fregoso2019}.

\begin{figure}[]
\subfigure{\includegraphics[width=.45\textwidth]{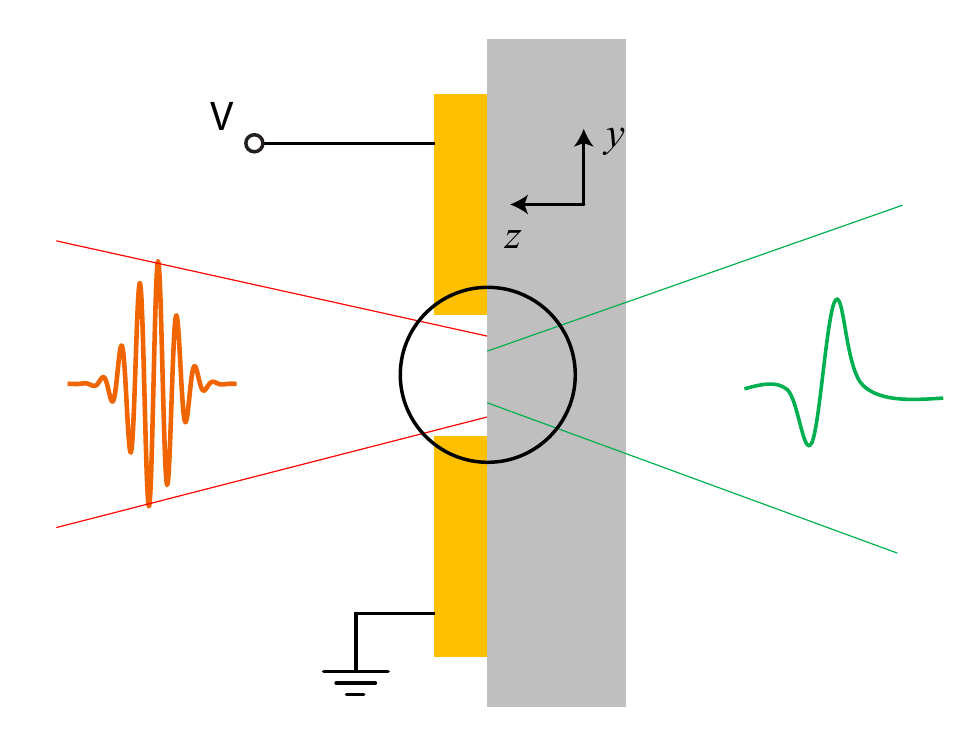}}
\caption{The photoconductive switch shown in profile (adapted from Ref.~\onlinecite{Peiponen2013}). An incident pulse with central frequency above the energy band gap (red) produces a single-cycle pulse in the THz frequency regime (green).}
\label{fig:photoswitch} 
\end{figure}

Photoconductvity has long been studied in the context of generation of intense terahertz (THz) pulses in photoconductive dipole switches~\cite{Auston1984,Fattinger1988}. A typical photoconductive dipole switch consists of two metal electrodes on a semiconductor substrate (usually low-temperature GaAs) separated by a distance ranging from micrometers to centimeters (see Fig.~\ref{fig:photoswitch}). A potential difference between the electrodes establishes a static electric field $\v{E}_0$. An electric field pulse of femtosecond duration is incident on the gap between the electrodes and creates free charges which are then accelerated. The generated photocurrent radiates an electric field in the THz frequency range as\cite{Jackson1999}  
\begin{align}
\v{E}_{thz} \propto \frac{d \v{J}}{dt},
\label{eq:E_thz}
\end{align}
far from the source. Usually, the pulse has a gaussian shape with a central frequency just above the energy band gap. The carriers are  often assumed to form a free-electron gas with the total number of carriers available for conduction being proportional to the intensity of the pulse envelop. The dynamics of the driven gas has been studied extensively using isotropic Drude-like models of conduction with phenomenological parameters involving  relaxation times, mobilities, etc. \cite{Benicewicz1994,Jepsen1996,Duvillaret2001}. Single-cycle THz pulses from photoconductive switches find many applications in materials science, medicine, biology and the military, and hence it is important to develop an understanding of how the crystal symmetry affects the magnitude and polarization of the emitted THz field. Such effects are not captured by isotropic models of conduction. For a review of the vast literature on the subject, we refer the reader to the many excellent reviews available~\cite{Burford2017,Dhillon2017,Peiponen2013}.

\subsection{Jerk current}
Consider an optical field $\v{E}$ and a static field $\v{E}_0$. An expansion of the photoconductivity in powers of these fields gives three contributions\cite{Fregoso2018,Fregoso2019} which we write as
\begin{align}
\v{J}_{dc,ph}^{(3)}= \iota_{3} \v{E}^2 \v{E}_0 +  \eta_{3} (\v{E}\times\v{E}^{*}) \v{E}_0 +   \sigma_{3} \v{E}^2 \v{E}_0.
\label{eq:ph}
\end{align}
The first term is the so-called jerk current and it is expected to be the largest contribution to photoconductivity. The second and third terms are higher-order versions of the injection and shift currents of the BPVE. For a monochromatic and spatially homogeneous field $E^a\equiv E^a_{\omega} e^{-i\omega t} + c.c.$, the jerk current obeys the phenomenological equation (see Appendix~\ref{eq:relaxation_pheno})
\begin{align}
\frac{d^2}{dt^2} J^{a(3)}_{jerk} = 6 \iota_{3}^{abcd}(0,\omega,-\omega,0) E^b_{\omega} E^c_{-\omega} E^d_0 + \frac{1}{\tau^2}J^{a(3)}_{jerk},
\label{eq:Jacc}
\end{align}
where summation over repeted indices is implied and $\iota^{abcd}_3(0,\omega,-\omega,0)$ is given by 
\begin{align}
\iota_3^{abcd} = \frac{2\pi e^4}{6\hbar^3 V}\sum_{nm\v{k}} &f_{mn}  \omega_{nm;ad} r^b_{nm}r^c_{mn}\delta(\omega_{nm}-\omega).
\label{eq:jerk}
\end{align}
Here $a,b,c,d=x,y,z$ are Cartesian components, $n,m$ are band indices, $\v{k}$ is the crystal momentum in the Brillouin zone (BZ), $f_{nm}\equiv f_n-f_m$ is the difference in occupation numbers at zero temperature of bands $n$ and $m$, $\hbar\omega_n$ is the energy of the band $n$, $\omega_{nm}\equiv\omega_n - \omega_m$ band energy differences, $r^{a}_{nm}=i\langle u_{n}|\partial u_{m}/\partial k^a \rangle$ is the the Berry connection, $u_{n}$ is the periodic part of the Bloch wave function, $\omega_{nm;ad}\equiv \partial^2 \omega_{nm}/\partial k^d \partial k^a= \partial^2 \omega_{n}/\partial k^d \partial k^a-\partial^2 \omega_{m}/\partial k^d \partial k^a $, and $e$ is the signed electron charge. We assumed a simple phenomenological model of relaxation with a single relaxation time scale $\tau$. The jerk current tensor is symmetric under exchanges of $b \leftrightarrow c$ and $a \leftrightarrow d$ and is a measure of the curvature of the energy bands. 

The source term in Eq.~(\ref{eq:Jacc}) admits a simple semiclassical interpretation. To see this, take two time derivatives of 
\begin{align}
\v{J}=\frac{e}{V} \sum_{n\v{k}} f_n \v{v}_n,
\label{eq:current}
\end{align}
where $\v{v}_n(\v{k})$ is the Bloch velocity in Bloch state $n$,$\v{k}$. Keeping terms to $\mathcal{O}(\v{E}^2 \v{E}_0)$ we have
\begin{align}
\frac{d^2}{dt^2} \v{J}_{dc,jerk}^{(3)} = \frac{e}{V} \sum_{n\v{k}}\bigg[2 \frac{d f_n}{dt}^{(2)}\frac{d\v{v}_n}{dt}^{(1)} + \frac{d^2 f_n}{dt^2}^{(3)} \v{v}_n\bigg].
\label{eq:jerk_intuitive}
\end{align}
We left out a term which contributes in metals that break time reversal symmetry~\cite{Matsyshyn2019}. Superscripts $(n)$ indicate the $nth$-order in the electric fields. The first term represents a process where carriers pumped into conduction states accelerate uniformly with $d v_n^a/dt=e\omega_{n;ad} E^d_0/\hbar$ under the action of a static electric field. The carrier injection rate $df_n/dt$ is $\mathcal{O}(\v{E}^2)$ to lowest order (see Appendix~\ref{eq:relaxation_pheno}). 

The second term represents a process where the carrier injection rate accelerates. A simple cartoon picture is as follows. Since the wave vector of the wave packet evolves as $e\v{E}_0 t/\hbar$ in the presence of the static field, the time-reversed states $\pm \v{k}$ are Doppler shifted to $\pm \v{k}+e\v{E}_0 t/\hbar$ in the frame of reference where a wave packet is at rest. In this frame 
\begin{align}
\frac{d^2} {dt^2}f_n(-\v{k})  \neq \frac{d^2}{dt^2}f_n(\v{k}),
\end{align}
giving rise to a net current. The process is very different from light-matter-induced injection current processes where the velocity if carrier injection, not its acceleration, is asymmetric and only for circular or elliptic polarization. The jerk current, on the other hand, can be finite for both polarizations.  As shown in appendix \ref{eq:relaxation_pheno}, the second term in Eq.~(\ref{eq:jerk_intuitive}) is proportional to the first term and together they give Eq.~(\ref{eq:jerk}).

\section{Jerk current tensor in GaAs and monolayer GeS}
\begin{figure}[]
\subfigure{\includegraphics[width=.43\textwidth]{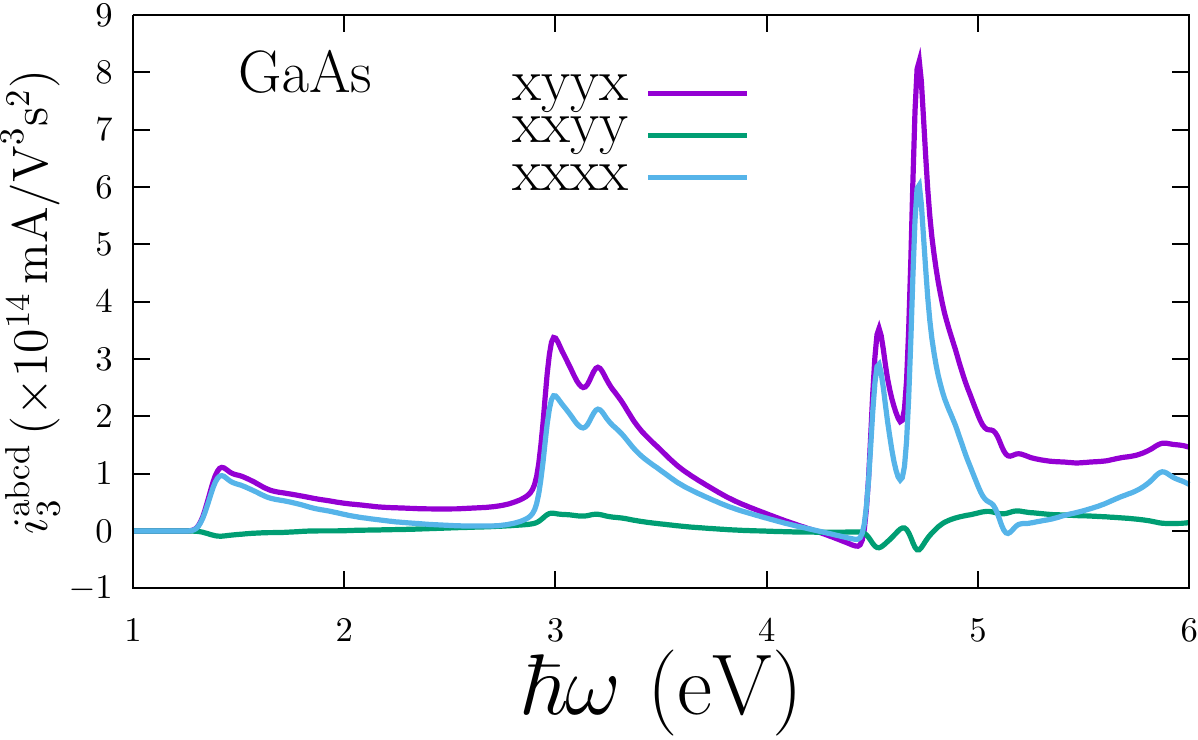}}
\caption{Independent components of $\iota_3^{abcd}$ in GaAs as a function of photon energy. The spectrum is composed of isolated peaks which coincide with a high joint density of states (see Appendix \ref{sec:jerk_gaas}.}
\label{fig:iota3_gaas}
\end{figure}

GaAs point group, $\bar{4}3m$, allows three independent components of $\iota_3^{abcd}$ shown in Fig.~\ref{fig:iota3_gaas} as a function of incoming photon energy. The numerical details of the density functional theory (DFT) calculation are presented in Appendix \ref{sec:methods}. The spectrum of the response tensor vanishes for photon energies lower than the energy band gap of 1.4 eV. The spectrum peaks at 1.4 eV, 3 eV and at 4.7 eV where it reaches 1$\times 10^{14}$ mA/V$^3$s$^{2}$, 3$\times 10^{14}$ mA/V$^3$s$^{2}$ and 6$\times 10^{14}$ mA/V$^3$s$^{2}$ respectively. Note that current transverse to the static field, controlled by the $xxyy$ component, is about an order of magnitude smaller with respect to the longitudinal components. As shown in the Appendix \ref{sec:jerk_gaas}, the isolated peaks in $\iota_3$ can be explained, in part, by a high joint density of states (JDOS) at various points in the BZ. The response tensor for Si has a similar spectrum (see Appendix \ref{app:jerk_si}).

\begin{figure}[]
\subfigure{\includegraphics[width=.43\textwidth]{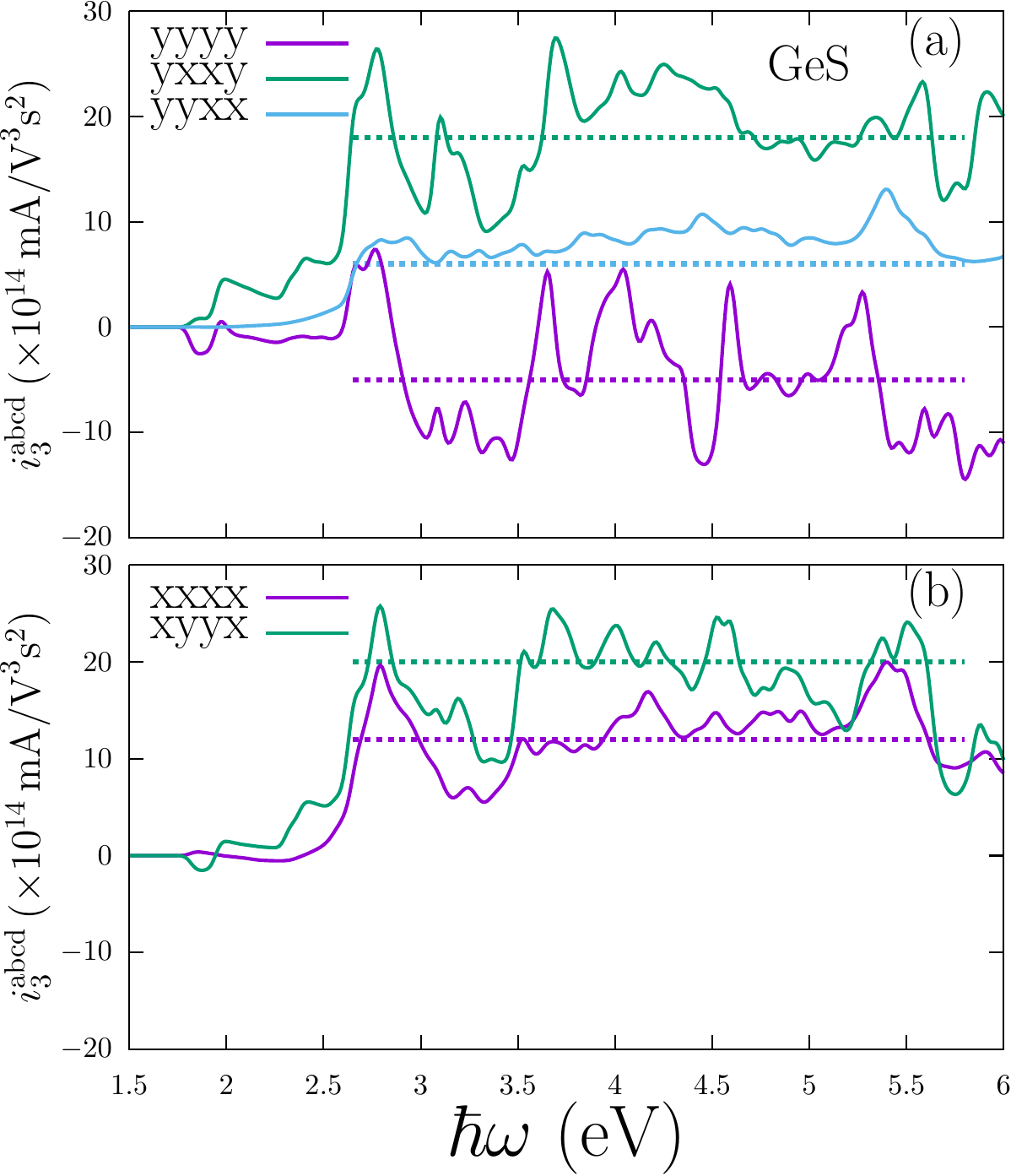}}
\caption{(a) and (b) Independent components of $\iota_3^{abcd}$ in monolayer GeS. The spectrum is flatter than in GaAs but has peak values of an order of magnitude larger. The spectrum of monolayer GeSe, SnSe, and SnSe is qualitatively similar and is presented in Appendix \ref{sec:jerk_ges}.}
\label{fig:iota3_ges}
\end{figure}

It is interesting to compare the spectrum of $\iota_3$ of GaAs which breaks inversion symmetry (but is not ferroelectric) with that of monolayer GeS (predicted to have large in-plane spontaneous polarization,\cite{Fei2016,Rangel2017}) and that of Si (which does not break inversion symmetry). Let us choose the slab to define the $xy$-plane with the $x$-axis along the polarization axis and $z$ out of the slab. The point group of monolayer GeS $mm2$ allows five (in-plane) independent components, shown in Fig.~\ref{fig:iota3_ges}. The out-of-plane response is much weaker and is not considered here. The 2D values are reported as bulk equivalent, i.e., per unit volume, to enable easy comparison to other materials. The spectrum of monolayer GeSe, SnS and SnSe are similar (see Appendix \ref{sec:jerk_ges}). 

$\iota_3$ in monolayer GeS is zero for frequencies smaller than the 1.9 eV energy gap, as expected. In contrast to GaAs and Si, the response is flatter over the 2.5-6 eV range of photon energies, which includes part of the visible spectrum. Importantly, the peak responses are about an order of magnitude larger than in GaAs and Si, and can reach peak values of $50 \times 10^{14}$ mA/V$^3$s$^{2}$ in monolayer SnSe (see Appendix \ref{sec:jerk_ges}). The transverse component $xxyy(=yyxx)$ are an order of magnitude smaller than the longitudinal components. Analysis of the JDOS, spin-orbit coupling (SOC), and polarization suggest that the reduced dimensionality and the symmetries imposed by the in-plane polarization are responsible for the larger response. A similar conclusion was reached in the case of injection current in these materials~\cite{Panday2019}.

\section{THz radiation of jerk current}
We now demonstrate that the plane of polarization of emitted THz pulses is rotated with respect to the direction of the bias field. This is in contrast to isotropic models of conduction which predict the polarization of emitted THz pulses to be parallel to the static field. Consider the width $w$ of the envelop to be much longer than the period the central frequency $T$. In such case, many cycles occur at an approximately constant amplitude. We also assume that the momentum relaxation time $\tau$ is small compared with $w$ but larger if compared to $T$, 
\begin{align}
w \gg \tau \gg T.
\label{eq:tau_pulse_width}
\end{align}
Under this conditions, it is easy to see that the magnitude of the instantaneous (time-varying) photocurrent is proportional to the (time-varying) intensity of the envelop times the jerk current tensor evaluated at the central frequency ~\ref{eq:jerk}. In practice, these assumptions are not so stringent, since for typical clean semiconductors with an energy band band gap in the visible, $T\sim 2$ fs, $\tau\sim 50$ fs and $w\sim 200$ fs (see for example Refs.\cite{Cote2002,Bieler2005,Laman2005}). In addition, we assume that the injected carriers do not saturate the semiconductor. This is usually the case for moderate (but experimentally relevant) bias fields and optical powers~\cite{Benicewicz1994}. The advantages of working in this regime are as follows. (i) The frequency dependance of $\iota_3$ provides useful information about the magnitude of the photocurrent response and hence of the THz field amplitude. For example, our calculation of $\iota_3$ above suggests that photon energies of 3 eV produce twice as much (jerk) photocurrent in GaAs than at the band edge. Previous experiments almost always use photon energies just above the band edge~\cite{Peiponen2013}. More important, monolayer GeS produces an order of magnitude larger (jerk) photocurrent than GaAs and Si. (ii) Lattice anisotropy could give rise to THz radiation not along the bias field, an effect not captured in Drude-like models where the emitted THz field is parallel to $\v{E}_0$. 

\begin{figure}[]
\subfigure{\includegraphics[width=.43\textwidth]{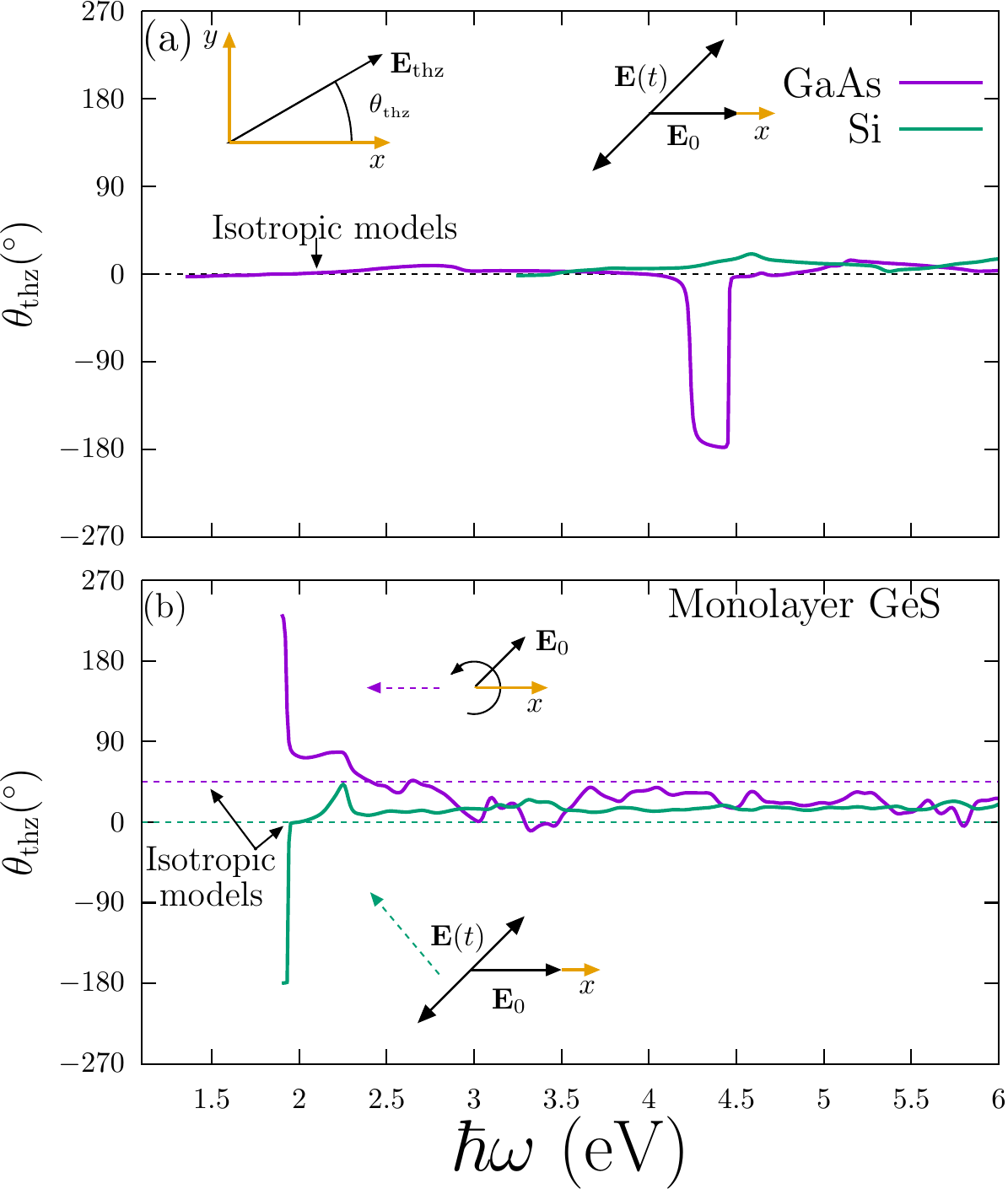}}
\caption{(a) Angle of polarization of emitted THz pulses in a GaAs and Si Hertzian dipoles. The optical field is polarized at 45$^\circ$ and the static field $\v{E}_0=\hat{\v{x}}E_0^x$. Isotropic models predict a response parallel to $\v{E}_0$ (dashed lines). The angle of the emitted THz pulse does not depend on the magnitude of the static field or the optical field, nor on the details of the relaxation mechanism. (b) Same as (a) but for monolayer GeS. Two polarizations of light are indicated along with the isotropic model predictions (dashed lines). For circular polarization, the static field is at 45$^\circ$. For linear polarization, the optical field is at 45$^\circ$ and $\v{E}_0=\hat{\v{x}}E_0^x$. }
\label{fig:theta_thz_gaas}
\end{figure}

As a first example, consider an incident optical field $\v{E}$ perpendicular to the $\v{ab}$-plane of GaAs assumed to coincide with the lab's $xy$-plane (see Fig.~\ref{fig:photoswitch}). There is also a static in-plane field $\v{E}_0$ 
\begin{align}
\v{E} &= \hat{\v{x}} E^{x}_{\omega} e^{-i\omega t} + \hat{\v{y}} E^{y}_{\omega} e^{-i\omega t} + c.c.,\\
\v{E}_0 &= \hat{\v{x}} E^{x}_{0} + \hat{\v{y}} E^{y}_{0},
\end{align}
where $E^{x}_{\omega} = |E^{x}_{\omega}|e^{-i\phi_x }$, $E^{y}_{\omega} = |E^{y}_{\omega}|e^{-i\phi_y }$. The angle of the photocurrent defines the angle of the THz pulse polarization, see Eq.~(\ref{eq:E_thz}). Let us define $\theta_{thz}$ with respect to the $x$-axis as 
\begin{align}
\tan\theta_{thz} = \frac{J^y_{jerk}}{J^x_{jerk}}.
\label{eq:theta_thz_def}
\end{align}
An explicit calculation gives
\begin{align}
\theta_{thz} &= \theta_{\v{E}_0}. ~~~~~~~~~~~~~~~\textrm{(circ. pol.)}, \\
\tan\theta_{thz} &= \frac{2 \iota_3^{xxyy}}{\iota_3^{xxxx} + \iota_3^{xyyx}} ~~\textrm{(45$^\circ$ pol. $\v{E}_0 \parallel \hat{\v{x}} $)},
\end{align}
where $\theta_{\v{E}_0}$ is the angle of the bias field. The first result agrees with the predictions of isotropic models~\cite{Peiponen2013}. The second result, obtained for a linearly polarized optical field at an angle of 45$^\circ$ and $\v{E}_0=\hat{\v{x}}E^x_0$, is different from $\theta_{thz}=0$ predicted by isotropic models. Note that the angle is independent of the strength of the static and optical fields and of the microscopic parameter $\tau$; it is an intrinsic property of the Bloch wave function. The amplitude of the emitted THz field is proportional to the intensity of the pulse envelop and the bias field. Naturally, the power emitted is proportional to the square of the bias field, since $\v{E}_{thz}^2\sim \v{E}_{0}^2$, which is in agreement with the scaling analysis of experimental results~\cite{Benicewicz1994}. In Fig.~\ref{fig:theta_thz_gaas}(a) we show $\theta_{thz}$ for GaAs and Si. The small angle at frequencies near the band edge is consistent with the lack of anisotropy found in experiments where frequency was adjusted to just above the band edge~\cite{Peiponen2013}. Note that injection current  vanishes in GaAs by symmetry, and shift current, which can be generated, vanishes for the geometry we consider. Optical rectification-based THz pulses can be shown to be smaller than jerk-based THz radiation for long pulse optical excitation.

Now consider the angle of rotation of the polarization of THz pulses in monolayer GeS. Let the polar axis define the $x$-axis and the plane of the slab define the $xy$-plane (see Fig.~\ref{fig:photoswitch}). The angle of rotation for two representative cases is 
\begin{align}
\tan\theta_{thz} &= \frac{\iota_3^{yxxy} + \iota_3^{yyyy}}{\iota_3^{xxxx} + \iota_3^{xyyx}}\tan\theta_{\v{E}_0} \textrm{~~(circ. pol.)}, \\
\tan\theta_{thz} &= \frac{2 \iota_3^{yyxx}}{\iota_3^{xxxx} + \iota_3^{xyyx}} ~~~~\textrm{(45$^\circ$ pol. $\v{E}_0\parallel \hat{\v{x}} $)},
\end{align}
and is shown in Fig.~\ref{fig:theta_thz_gaas}(b). The angles differ from the predictions of isotropic models. In fact, the polarization can turn almost 180$^\circ$ for frequencies near the band edge in the visible regime. Since monolayer GeS breaks inversion symmetry, injection and shift current will be generated. The injection current contributes $J^y_{inj} = 4i\tau \eta_2^{yyx}|E^y_{\omega}|^2$ to circular polarization~\cite{Panday2019}. In typical experimental setups, the static field $10^5$ V/m will generate a jerk current larger than the injection current. Shift current contributes to linear polarization\~cite{Rangel2017} with an angle $\tan\theta_{thz,shift} = 2\sigma_2^{yyx}/(\sigma_2^{xxx}+\sigma_2^{xyy})$, but for typical static fields, jerk current is expected to be much larger. An estimate of the relative magnitudes of these currents was given in Ref.\cite{Fregoso2019}

\section{Conclusion}
In this work we use DFT to compute the jerk current response tensor $\iota_3^{abcd}$ of prototypical semiconductors GaAs and Si and of  ferroelectric single-layer GeS, GeSe, SnS and SnSe as a function of photon energy. The jerk current spectrum of GaAs and Si is composed of isolated peaks largely explained by high JDOS, and which can reach values up to 3$\times 10^{14}$ mA/V$^3$s$^2$ within the visible spectrum. The spectrum of monolayer GeS, GeSe, SnS and SnSe, on the other hand, is flatter over a range of photon energies which includes the visible regime and peaks at values about an order of magnitude larger than those in GaAs and Si. 

The difference with previous approaches is that have incorporated the effects of the crystal symmetry into the acceleration process of charges and have shown that the detailed knowledge of $\iota_3^{abcd}$ makes it possible to predict the frequency at which THz output power is largest. This provides a possible solution to the main drawback of Hertzian dipole technology, namely, degradation of emitters due to high bias voltage, subsequent thermal load, and saturation effects. In addition, a large rotation of the polarization plane of the emitted THz pulse is possible and could be used to produce on-demand THz pulses with a given polarization by either changing the bias voltage or the frequency, or both. Finally, jerk current does not require crystals without inversion symmetry, and hence, a wide range materials could be explored for applications.

\section{Acknowledgments}
We thank the authors of Ref.~\onlinecite{Ventura} for alerting us to some errors in earlier versions of the paper. We also thank DOE-NERSC contract No. DE-AC02-05CH11231 and CONACYT project A1-S-9410 for their support. 

\appendix

\begin{table*}
\caption{Peak values of $|\iota^{abcd}_3|$ in various semiconductors obtained from DFT. The photon energy, direct experimental gap, and spontaneous polarization $|\v{P}_0|=P_0$ are indicated. For 2D materials the bulk equivalent value is reported.} 
\begin{center}
 \begin{tabular}{|l|c|c|c|c|c|}
 \hline 
  ~~~~~~Material& ~ $|\iota^{abcd}_3|$~ ($\times 10^{14}$ mA/V$^3$s$^2$) ~~~ & ~~$\hbar\omega$~ (eV)~~ & ~~Direct gap (eV)~~ & $P_0$~ ($\mu$C/cm$^2$) & ~~Inversion symmetry~~ \\
 \hline 
  ~~Monolayer SnSe~~    & 50                           & 3.1             &  0.95  & 0.72   & no  \\
 \hline 
 ~~Monolayer GeSe~~     & 40                           & 2.5             &  1.16  &  1.38  & no \\ 
  \hline 
 ~~Monolayer GeS~~      & 30                           & 2.7             &  1.89  & 1.95   & no \\
\hline 
 ~~Monolayer SnS~~      & 30                           & 3.0             &  1.57  &  0.95  & no \\
 \hline 
  ~~~~~~Si     				  & 12                           & 3.5             &  3.4   &  0     & yes \\
 \hline 
  ~~~~~GaAs     			  & 3                            & 3               &  1.42  &  0     & no \\
 \hline 
\end{tabular}
\end{center}
\label{table:iota3}
\end{table*}

\section{Phenomenology of jerk current}
\label{eq:relaxation_pheno}
The calculation of $\iota_3$ is based on divergences of susceptibilities.\cite{Fregoso2018,Fregoso2019} The response tensor, however, also admits a simple semiclassical interpretation which helps develop physical intuition. Let us take two time derivatives of the current
\begin{align}
\v{J}=\frac{e}{V} \sum_{n\v{k}} f_n \v{v}_n,
\label{eq:current_app}
\end{align}
and only consider processes of order $\mathcal{O}(\v{E}^2\v{E}_0)$ to obtain 
\begin{align}
\frac{d^2}{dt^2} \v{J}_{dc,jerk}^{(3)} = \frac{e}{V} \sum_{n\v{k}}\bigg[2 \frac{d f_n}{dt}^{(2)}\frac{d\v{v}_n}{dt}^{(1)} + \frac{d^2 f_n}{dt^2}^{(3)} \v{v}_n\bigg]. 
\label{eq:jerk_intuitive_v2}
\end{align}
The superscripts indicate the order in the electric field. The factors in the first term are
\begin{align}
\frac{d v^a_n}{dt} &=\frac{e}{\hbar}\omega_{n;ad} E^d_0,  
\label{eq:dvdt} \\
\frac{d f_c}{dt} &= \frac{2\pi e^2}{\hbar^2}\sum_v |\v{E}_{\omega}\cdot \v{r}_{cv}|^2 \delta(\omega_{cv}-\omega), 
\label{eq:FGR1}\\
\frac{d f_v}{dt} &= -\frac{2\pi e^2}{\hbar^2}\sum_c |\v{E}_{\omega}\cdot \v{r}_{cv}|^2 \delta(\omega_{cv}-\omega).
\label{eq:FGR2} 
\end{align}
In (\ref{eq:dvdt}) we used $v^a_n =\omega_{n;a}$ and the semiclassical expression $d\v{k}/ dt= e\v{E}_0/\hbar$ for the temporal 
evolution of a wave vector of the wave packet under the action of the static field. In (\ref{eq:FGR1}) the rate of carrier injection is given by Fermi's golden rule. Substituting (\ref{eq:dvdt}) to (\ref{eq:FGR2}) into Eq.~(\ref{eq:jerk_intuitive_v2}) we recover Eq.~(\ref{eq:jerk}) in the main text. 

We can incorporate momentum relaxation phenomenologically by adding terms $-v^a_n/\tau_m$, $-(f_n-f^{(0)}_n)/\tau_r$ and incorporate space charges effects, etc. Since we are interested only in anisotropic effects at short time scales, substitution of the above dissipative terms gives the same basic relaxation model already captured in Eq.~(\ref{eq:Jacc}) and couplings between the jerk current and the injection current which are assumed small in our experimental setup.  

The second term in Eq.~(\ref{eq:jerk_intuitive_v2}) represents a shift in carrier injection rates due to the relative motion of wave packets (a Doppler shift) as explained in the main text. We can also see this in a simpler way as follows. Take a derivative of Eq.~(\ref{eq:FGR1}) to obtain
\begin{align}
\frac{d^2 f_c}{dt^2} &= \frac{2\pi e^3}{\hbar^3}\sum_v  \frac{\partial (r^a_{cv} r^b_{vc})}{\partial k^d}   E^a_{\omega} E^b_{-\omega}E^d_0 \delta(\omega_{cv}-\omega)\nn \\
&+ \frac{2\pi e^3}{\hbar^3}\sum_v  (r^a_{cv} r^b_{vc})   E^a_{\omega} E^b_{-\omega}E^d_0 \frac{\partial }{\partial k^d}\delta(\omega_{cv}-\omega)
\end{align}
from which it is clear that there is an asymmetry in the \textit{acceleration} of carrier injection at time reversed points in the BZ 
\begin{align}
\frac{d^2} {dt^2}f_{c}(-\v{k})  \neq \frac{d^2}{dt^2}f_{c}(\v{k})
\end{align}
for linearly and circularly polarized fields. Let us compare this process with light-induced injection current processes. In injection processes we take a time derivative of Eq.~(\ref{eq:current_app}) and keep only terms second order 
\begin{align}
\frac{d}{dt} \v{J}^{(2)}_{dc,inj} = \frac{e}{V} \sum_{n\v{k}} \frac{d f_n}{dt}^{(2)}\v{v}_n.
\label{eq:inj_intuitive}
\end{align}
By Eq.~(\ref{eq:FGR1}) or (\ref{eq:FGR2}) it is easy to show that there is an asymmetry in the \textit{velocity} of carrier injection 
\begin{align}
\frac{d}{dt}f_c(-\v{k}) \neq \frac{d}{dt}f_c(\v{k}),
\end{align}
when the field is complex, leading to a net (injection) current. At times much longer than the period of the optical field but smaller than the effective relaxation time, $\iota^{abcd}_3$ acts as a current source of magnitude
\begin{align}
J^{a(3)}_{jerk}= 6\tau^2 \iota_{3}^{abcd}(0,\omega,-\omega,0) E^b_{\omega} E^c_{-\omega} E^d_0,
\end{align}
and its time dependance follows the pulse envelope.

\section{Numerical methods}
\label{sec:methods}
We use DFT as implemented in the ABINIT~\cite{Gonze2009} computer package with the generalized gradient approximation to the exchange correlation energy functional as implemented by Perdew \textit{et al.}~\cite{Perdew1996}. Hartwigsen-Goedecker-Hutter norm conserving pseudo potentials~\cite{Hartwigsen1998} were employed. To expand the plane waves basis set, energy cutoffs of 50 Hartree were employed for GaAs, Si and monolayer GeS and GeSe and 60 Hartree for SnS and SnSe. We chose the plane of the slab to define the $xy$-plane with the $x$-axis along the spontaneous polarization. The lattice parameter in the $z$-direction was set to 15 \AA, which makes for more than 10 \AA~ of vacuum between slabs. To calculate $\iota^{abcd}_3$, we included 20 valence and 30 conduction bands for GaAs, Si and GeS and SnS, and 30 valence and 20 conduction bands for GeSe and SnSe. They account for all allowed transitions up to 6 eV.  

To extract the effective response of a single layer, we scaled the numerical result by factor $\mathcal{L}/d$, where $\mathcal{L}$ is the supercell lattice parameter perpendicular to the slab and $d$ is the effective thickness of the monolayer. For concreteness, we
estimated  slab thicknesses as 2.56, 2.59, 2.85 and 2.76 \AA~ for GeS, GeSe, SnS, and SnSe, respectively.  Once the ground-state wave
function and energies were computed, the TINIBA package~\cite{tiniba} was used to compute $\iota^{abcd}_3$ as implemented in Ref.\cite{Fregoso2018,Fregoso2019}. The sum over $\v{k}$-points uses the interpolation tetrahedron method\cite{bloch-tetra} as implemented in TINIBA.

%%%%%%%%%%%%%%%%%%%%%%%%%%%%%%%%%%%%%%%%%%%%%%%%%%%%%%%%%%%%%%%%%%%%%%%%%%%%%
\section{Jerk current tensor in GaAs}
\label{sec:jerk_gaas}
\begin{figure}[]
\subfigure{\includegraphics[width=0.45\textwidth]{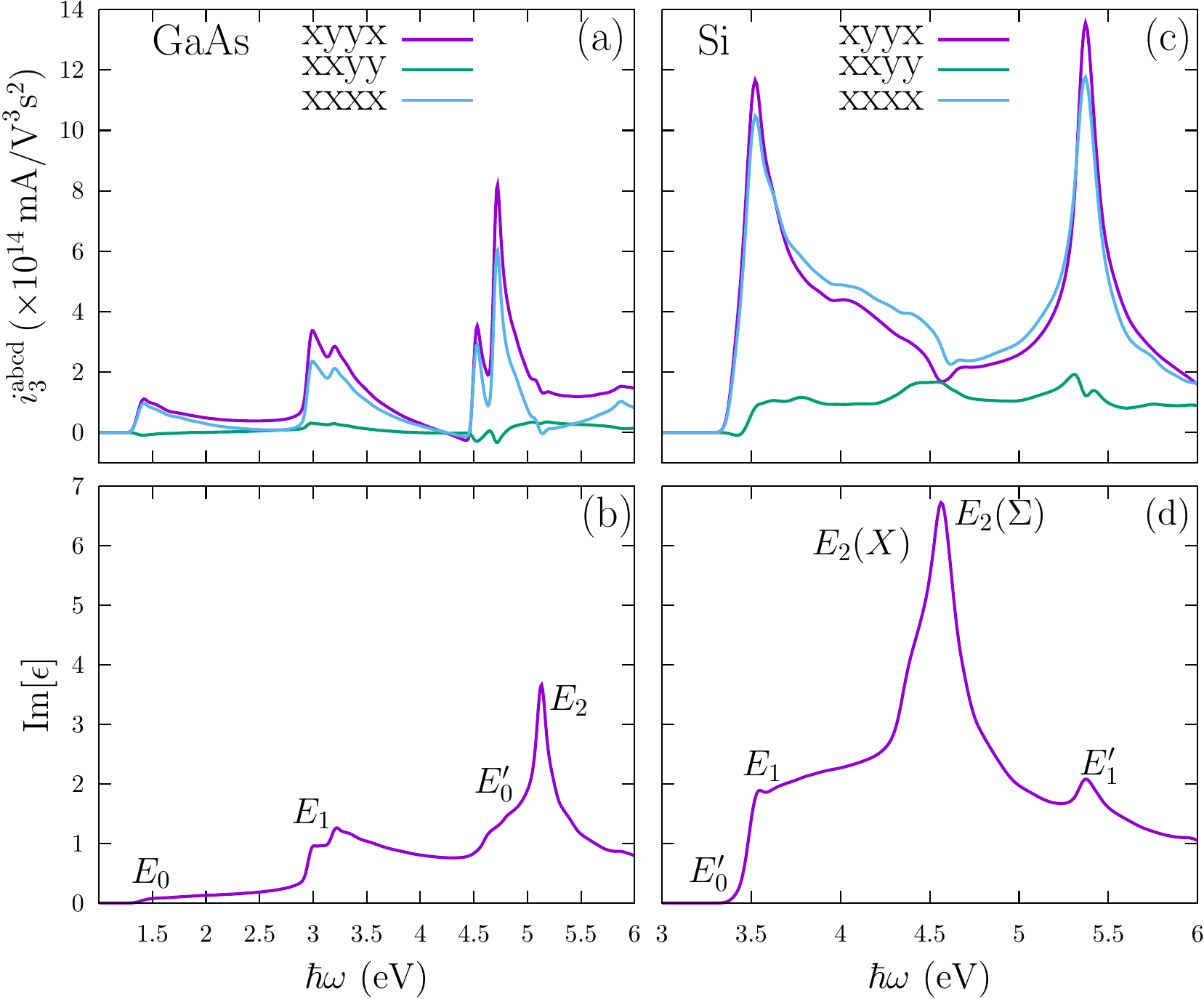}}
\caption{(a) Independent components $\iota_3^{abcd}$ in GaAs as a function of photon energy and (b) imaginary part of the dielectric function showing points with high JDOS $E_0$,$E'_0$, $E_1$, and $E_2$ in the BZ. (c) and (d) Same as (a) but for Si. The high JDOS can explain some of the peaks in $\iota^{abcd}_3$.} 
\label{fig:dos_vs_iota3_gaas}
\end{figure}
GaAs point group $\bar{4}3m$ allows three independent tensor components $\iota^{abcd}_3$, shown in Fig.~\ref{fig:dos_vs_iota3_gaas}(a) as a function of incoming photon energy.  

\subsection{Joint density of states}
It is interesting to compare the spectrum of jerk tensor with the spectrum of the linear imaginary dielectric function which roughly  follows the joint density of states (JDOS) and is a measure of the energy absorption by the material. The imaginary part of the linear dielectric function 
\begin{equation}
\textrm{Im}\varepsilon^{ab}(\omega)=\frac{e^{2}\pi}{\epsilon_{0}\hbar V}  \sum_{nm\v{k}} f_{nm} r^{a}_{nm}r^{b}_{mn}\delta{(\omega_{mn}-\omega)}, 
\label{eq:epsilon2}
\end{equation}
peaks at points in the BZ which have a high density of states. Four of these points, labeled by $E_0$, $E_1$, $E'_0$ and $E_2$, are shown in Fig.~\ref{fig:dos_vs_iota3_gaas}(b). As can be seen, the peak at 3 eV in all of the components of $\iota^{abcd}_3$ coincides with $E_1$. The peak at 4.5 eV correspond to a singular point. Here $E_2$ is seen to have only a small influence.

\begin{figure}[]
\subfigure{\includegraphics[width=.43\textwidth]{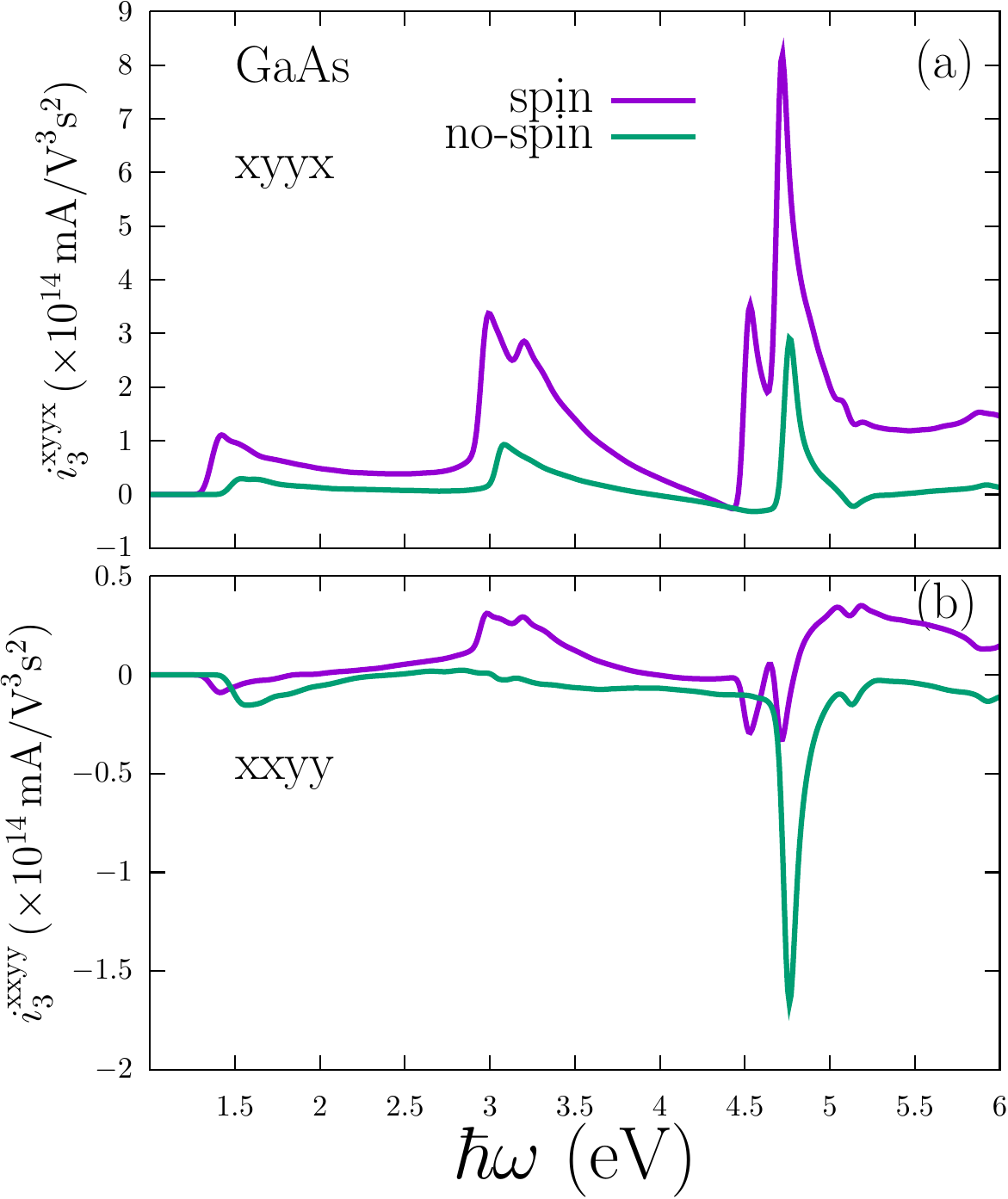}}
\caption{Comparison of $\iota^{abcd}_3$ in GaAs with and without SOC. Spin-orbit coupling splits the peaks into up and down components and shifts the response towards low frequencies.}
\label{fig:iota3_gaas_soc_vs_no_soc}
\end{figure}

\begin{figure*}[]
\subfigure{\includegraphics[width=1.0\textwidth]{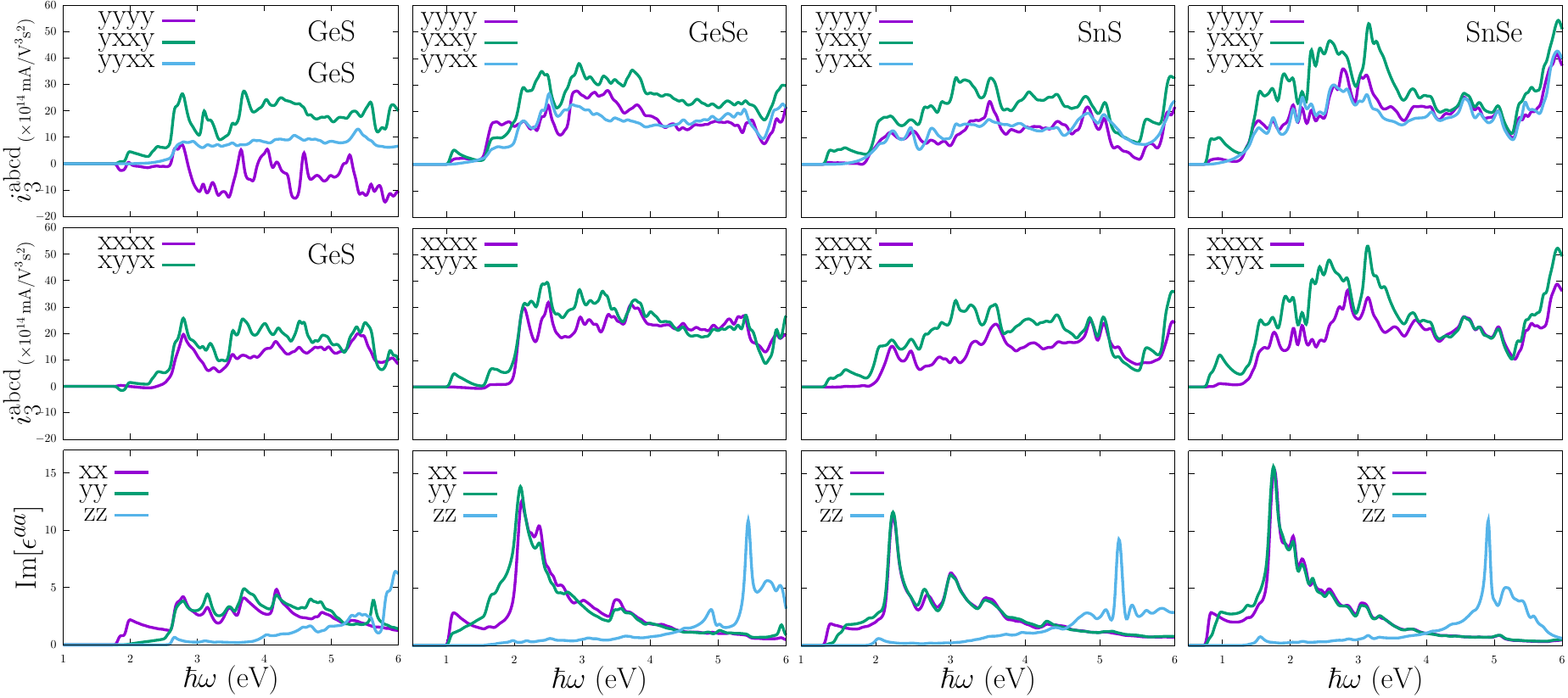}}
\caption{Top and middle rows show the independent components of $\iota_3^{abcd}$ in monolayer GeS, GeSe, SnS, and SnSe. The spectrum is flatter than the $\iota_3$-spectrum GaAs and Si. Note that peak values are about an order of magnitude larger. The bottom row shows the imaginary part of the linear dielectric function Eq.~(\ref{eq:epsilon2}) which indicates points of high JDOS and energy absorption. The correlation between to two responses is not as strong as in GaAs and Si.} 
\label{fig:iota3_ges_v2}
\end{figure*}

\subsection{Spin-orbit coupling}
Spin-orbit coupling (SOC) is known to be important in materials that break inversion symmetry such as GaAs. In GaAs, the SOC field is large near the band edge and hence we expect important corrections to the jerk tensor spectrum from SOC. In Fig.~\ref{fig:iota3_gaas_soc_vs_no_soc}(a) and \ref{fig:iota3_gaas_soc_vs_no_soc}(b) we show the $xyyx$ and $xxyy$ components of the $\iota^{abcd}_3$ with and without SOC. The main effect of SOC is to split and shift the spectrum towards lower energies. This is most noticeable near the points of high JDOS. The longitudinal component $xyyx$ increases significantly, whereas the transverse component decreases. 

\section{Jerk current in Si}
\label{app:jerk_si}
Si point group $m3m$ allows three independent tensor components $xyyx$, $xxyy$, and $xxxx$, shown in Fig.~\ref{fig:dos_vs_iota3_gaas}(c) and \ref{fig:dos_vs_iota3_gaas}(d) as a function of photon energy. The jerk spectrum vanishes for incoming photon energies lower than the direct energy band gap of 3.4 eV. It exhibits similar isolated peaks as GaAs. The spectrum peaks on the band edge and at 5.4 eV, both of the order of 12$\times 10^{14}$ mA/V$^3$s$^{2}$. Note that current transverse to the static field, controlled by the component $xxyy$ of $\iota^{abcd}_3$, is an order of magnitude smaller with respect to the longitudinal components. Similar to GaAs, peaks at 3.5 eV and 5.4 eV coincide with singularities in the JDOS.

%%%%%%%%%%%%%%%%%%%%%%%%%%%%%%%%%%%%%%%%%%%%%%%%%%%%%%%%%%%%%%%%%%%%%%%%%%%%%%%
\section{Jerk current tensor in monolayer GeS, GeSe, SnS and SnSe}
\label{sec:jerk_ges}
Monolayer GeS, GeSe, SnS and SnSe are predicted to have large in-plane spontaneous polarization (see Table~\ref{table:iota3}). Let us choose the slab to define the $xy$-plane with the $x$-axis along the polarization axis and $z$ pointing out of the slab. The point group of monolayer GeS is $mm2$ and accordingly there are six (in-plane) independent components  $yyyy$, $xxxx$, $yxxy$, $xyyx$, $yyxx$, and $xxyy$, shown in Fig.~\ref{fig:iota3_ges_v2} as a function of photon energy. The 2D values are reported as bulk equivalent, i.e., per unit volume, to easily compare to GaAs and Si. The out-of-plane response is much weaker and is not considered here. 
\begin{figure}
\subfigure{\includegraphics[width=.43\textwidth]{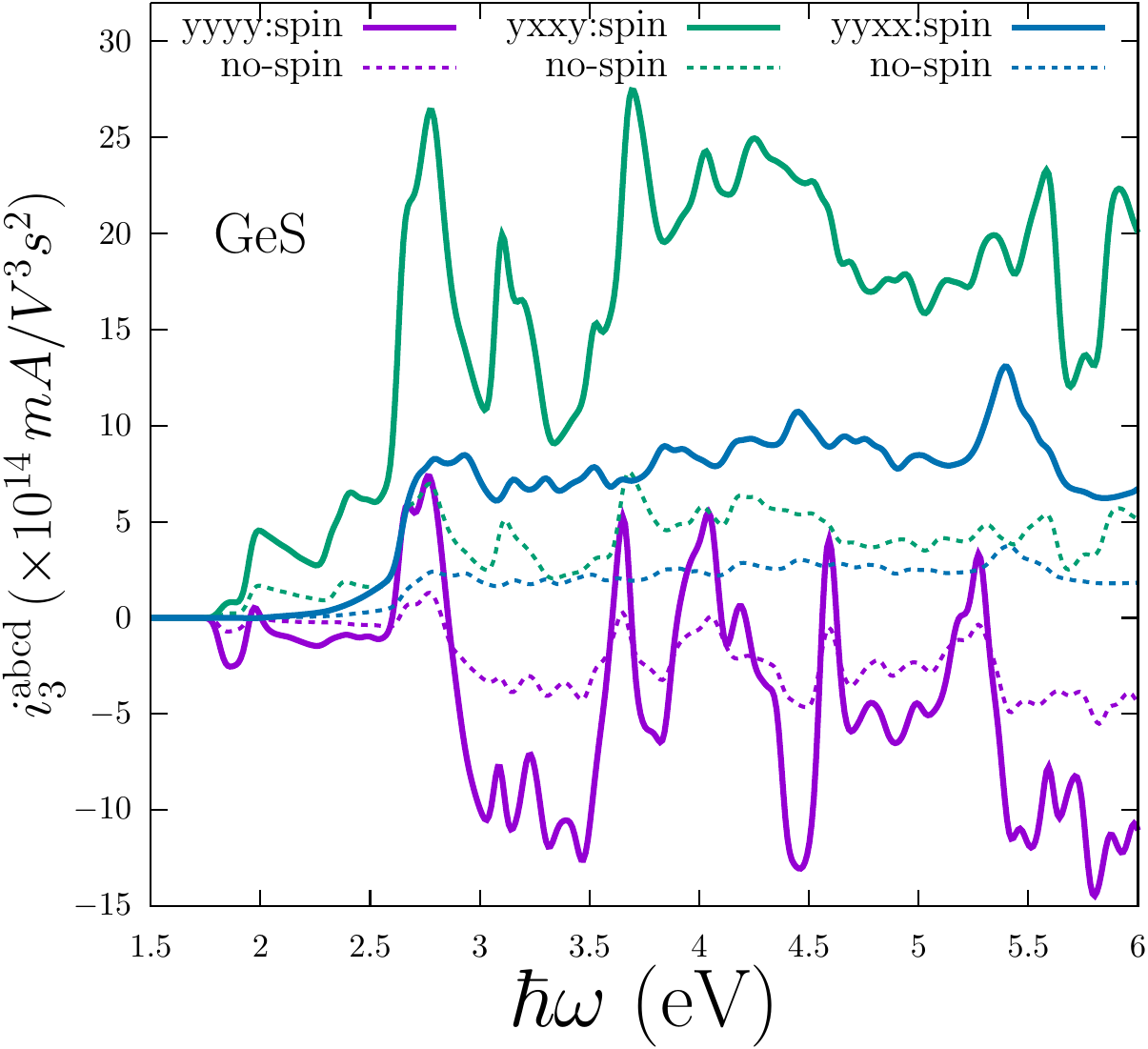}}
\caption{Longitudinal components of the jerk current tensor $yyyy$, $yxxy$ and transverse component $yyxx$, with and without SOC. Spin-orbit coupling enhances the response ten times for the longitudinal component, but not for the transverse component.} 
\label{fig:ges_iota3_soc_vs_nosoc}
\end{figure}
The spectrum of $\iota^{abcd}_3$ in monolayer GeS GeSe, SnS and SnSe is flatter than the spectrum of GaAs and Si in a broad range of photon energies (2.7-5.5 eV) including the visible spectrum. Peak values can reach $50\times 10^{14}$ mA/V$^3$s$^{2}$ in longitudinal components $yyyy$, $yxxy$, and $zxxz$. The transverse components $yyxx$ and $xxyy$ are in general an order of magnitude smaller when than the longitudinal components.  

Note the large difference in the response along and perpendicular to the polar axis. This is explained by the anisotropy of the lattice of these materials. Note also that the current perpendicular to the polarization described by $yyyy$ and $yxxy$ is very sensitive to the direction of the optical field, described by the middle two indices, specially in monolayer GeS. A change in the polarization of the optical field from $x$ to $y$ can change the sign of the photocurrent. This is not the case for the current along the polar axis described by $xxxx$, $xyyx$, where a change in optical field polarization from $x$ to $y$ will in general only decrease/increase the jerk current tensor.    

\subsection{Joint density of states} 
The JDOS of monolayer GeS (and related materials) has large fluctuations, see the bottom row in Fig.~\ref{fig:iota3_ges_v2}. The overall magnitude of the JDOS of monolayer GeS is at most twice that of GaAs, which suggests that the absolute magnitude of the JDOS does not explain the origin of the 10$\times$ difference in $\iota^{abcd}_{3}$ in our 2D materials. This suggest that other factors, such as reduced dimensionality, play a role. 

\subsection{Spin-orbit coupling} 
Spin-orbit coupling in monolayer GeS (and related materials) is small (of the order of $\sim 20$ mV), yet SOC increases the jerk spectrum by threefold with respect to the case with no SOC for all longitudinal components. Fig.~\ref{fig:ges_iota3_soc_vs_nosoc} shows the comparison of the jerk current spectrum for the largest components of monolayer GeS, $yyyy$, $yxxy$, and a transverse component $yyxx$. GaAs also shows an enhanced response of the longitudinal components with SOC, but by a lower factor, which suggests that reduced dimensionality plays a big role in the response of these 2D materials.

%\bibliography{iota3}

%merlin.mbs apsrev4-1.bst 2010-07-25 4.21a (PWD, AO, DPC) hacked
%Control: key (0)
%Control: author (8) initials jnrlst
%Control: editor formatted (1) identically to author
%Control: production of article title (-1) disabled
%Control: page (0) single
%Control: year (1) truncated
%Control: production of eprint (0) enabled
%

\end{document}